\documentstyle[12pt]{article}

\newcommand{\rf}[1]{(\ref{#1})}

\renewcommand{\thefootnote}{\fnsymbol{footnote}}

\newcommand{\newsection}{    % Numeration of eqs. is automatic
\setcounter{equation}{0}
\section}

\def\appendix#1{
  \addtocounter{section}{1}
  \setcounter{equation}{0}
  \renewcommand{\thesection}{\Alph{section}}
  \section*{Appendix \thesection\protect\indent \parbox[t]{11.715cm} {#1} }
  \addcontentsline{toc}{section}{Appendix \thesection\ \ \ #1}
  }

\def \v {{\rm v}}
\def \fff {{\rm  f}}

 \def \ep {\epsilon}
\def \L {{\Lambda}} 

\def \V {{\rm V}}
\def \r {{\rm r}}
\def \rr{{\bf r}}

\newcommand{\tr}{\ {\rm tr \ }}
\newcommand{\non}{\nonumber \\*}
\def \FFF{{\cal F}}
\def \ff {{\it f}}
\def \v{{\rm v}}

\def \f {{\bf f}}
\def\bea{\begin{eqnarray}}
\def\eea{\end{eqnarray}}
\def \ci {\cite}

\def\be{\begin{equation}}
\def\ee{\end{equation}}
\def\G{\Gamma}
\def \ha{{\textstyle{1\over 2}}}
\def \del {\partial}
\def \inv {^{-1}}
\def\Kt{{K'}}

\def\K{{\cal K}}
\def\W{{\cal W}}
\def \ov {\over }
\def\FFF{{\cal F}}
\def\F{{\bf F}}
\def\Q{{\cal Q}}
\def \e { e } 
\def \l {\lambda}
\def\T{{\rm T}}
\def \t{\theta}

\def\td{\tilde}
\def\a{\alpha}
\def \gym {g_{\rm YM}} 
\def\det{\hbox{det}}
\def \STr {{\rm STr }}
\def \SSTr{  {\widehat {\rm STr} } }

\def \Tr {{\rm Tr}}
\def\la{\label}
\def \foot{\footnote}

\def\np {{  Nucl. Phys. }}
\def \pl {{  Phys. Lett. }}

\def \pr  {{ Phys. Rev. }}

\def \bi{\bibitem}

\begin{document}
\begin{titlepage}
\begin{flushright}
ITP-SB-98-10\\
Imperial/TP/97-98/18\\
hep-th/9801120\\
\end{flushright}
\vspace{.5cm}

\begin{center}
{\LARGE  On Membrane Interaction in Matrix Theory }\\
\vspace{1.1cm}
{\large I. Chepelev${}^{{\rm 1,}}$\footnote{
E-mail: chepelev@insti.physics.sunysb.edu }
and A.A. Tseytlin${}^{{\rm 2,}}$\footnote{Also at Lebedev Physics
Institute, Moscow. \ E-mail: tseytlin@ic.ac.uk} }\\
\vspace{18pt}
${}^{{\rm 1\ }}${\it ITP, SUNY at Stony Brook, NY11794-3840}\\
${}^{{\rm 2\ }}${\it Blackett Laboratory, 
  Imperial College, London SW7 2BZ, U.K.}
\end{center}
\vskip 0.6 cm

\begin{abstract}
We compute the interaction potential between two parallel 
transversely boosted wrapped membranes (with fixed momentum $p_-$) 
in $D=11$ supergravity with compact light-like direction. We show 
that the supergravity result is in exact agreement with 
the potential following from the all-order Born-Infeld-type action
conjectured to be the leading planar infra-red part of the 
quantum super Yang-Mills effective action. This provides a non-trivial
test of consistency of the arguments relating Matrix theory
to a special limit of type II string theory. We also find 
 the potential between two (2+0) D-brane bound states in $D=10$
 supergravity (corresponding to the case of boosted membrane configuration 
in 11-dimensional theory compactified on a space-like direction).
We demonstrate that the result reduces to  the SYM expression
for the  potential in the special low-energy ($\a'\to 0$) limit,
in agreement with previous suggestions. 
In appendix we derive the action obtained from  the
$D=11$ membrane action by the world-volume duality transformation of 
the light-like coordinate $x^-$ into a 3-vector.
\end{abstract}

\end{titlepage}
\setcounter{page}{1}
\renewcommand{\thefootnote}{\arabic{footnote}}
\setcounter{footnote}{0}

%%%%%%%%%%%%%%%%%%%%%%%%%%%%%%%%%%
\newsection{Introduction}
%%%%%%%%%%%%%%%%%%%%%%%%%%%%%%%%%%

The  remarkable correspondence between  the 
super Yang-Mills (and thus matrix theory) and 
supergravity descriptions of interactions between   branes 
was originally  tested  in  the leading (one-loop) approximation
(see, e.g.,  \ci{dkps,bfss,AB,lifmat,lif3,CT1,CT2,KT}). 
 It was observed 
\ci{lifmat,CT1}  that to have   precise agreement between the 
interaction potentials derived from SYM and supergravity one
 needs to take  a  large $N$ 
(large  0-brane charge) limit in the  supergravity expression. 
In the simplest cases  this  can be effectively  accomplished by 
 subtracting the asymptotic value 1 from the harmonic function  $H$ in the action
 of a D-brane probe  moving  in the supergravity background
produced by a source brane.
As  was  observed  in \ci{bbpt} 
 on the example of subleading term in the interaction potential 
between   two D0-branes, 
 the required   supergravity expression 
can be obtained  automatically  by  keeping 
$N$ finite but considering   the $D=10$ configuration of 
branes    resulting  from an M-theory configuration 
 compactified on a light-cone direction 
 $x^- = x_{11} -t$ as in   the  finite $N$ proposal of \ci{suss}.\foot{The same supergravity potential is found  either by plugging the 
$x^-$-reduced background into  the D0-brane probe action in $D=10$
or considering the graviton probe action in $D=11$ and fixing the light-cone component of  momentum  $p_-$ \ci{bbpt}.} 
It  was checked  in \ci{CT3}, that a similar large $N$ or $H\to H-1$ recipe 
is important also for the  SYM-supergravity correspondence  at the level 
of subleading (two-loop) term in the interaction potential between 
a D-brane and a  BPS bound state of D-branes. 

A   suggestion   about  precisely which (`low-energy') limit of 
 $D=10$ supergravity should 
have  a  SYM description was  made in \ci{dps,maldacena}. 
A  related argument   providing a kinematical explanation 
for  the  correspondence   between  $D=11$  M-theory 
compactified on a light-like direction (with $p_-={N\ov R}$) 
 and  a  transverse p-torus  ($p \leq 3$) and  
a low energy, short distance, weak coupling  limit of a system of
$N$ Dp-branes in type II string theory on the dual p-torus 
 described by super Yang-Mills   theory 
was put forward in \ci{sen,seiberg}. 

According to \ci{dps,maldacena}, the tree-level supergravity 
describing configurations with RR charges
admits  a low-energy limit  in which it  {\it may} 
have a SYM description.
That limit does not formally require  taking   $N$ to be  large, 
but to   justify the possibility to ignore  closed string loop and 
$\a'$ corrections one needs also to assume that $g_s$ is small
and $g_s N$ is large.  As was observed in \ci{bbpt,seiberg}, 
the result of taking this  low-energy limit in $D=10$ supergravity 
expressions   should be 
 achieved automatically  by starting with
$D=11$ expressions and    taking the 
light-like direction  to be compact  (and fixing the value of $p_-$).

%Our results    support 
% the consistency of   the arguments of  \ci{maldacena,sen,seiberg}. 
The   observations made in
\ci{seiberg,sen} as such do  not  imply the agreement 
between  SYM (Matrix theory) 
 and supergravity \ci{hell}.  That agreement depends  on certain
special `non-renormalisation' properties of a class of planar  diagrams 
in string theory  and thus in 
maximally  supersymmetric  large N SYM theory \ci{dkps,bbpt,CT3}.

One aim of the present paper to  perform  a  test of the  formal 
arguments in \ci{maldacena} and \ci{seiberg} 
on the example of {\it all-order}  
 interaction  between  {\it extended}   BPS  branes.
Our results 
 provide also   another 
 test of  the Born-Infeld   ansatz \ci{CT3}
for the leading planar part of the perturbative SYM effective
 action which is central to this SYM-supergravity correspondence:
that  single  universal  expression  happens to describe 
  interaction potentials between various  types of branes
once one plugs in  appropriate  gauge field backgrounds.

We shall consider  a  potential between two parallel transverse 
$D=11$ membranes  having fixed values of the light-cone momentum.\foot{The non-perturbative SYM--supergravity
agreement  in the case of infinite 
 membrane scattering with $\Delta p_{11}\not=0$ was 
demonstrated (for the leading $O(v^4)$ term)  in \ci{PP,dkm}.
Related discussion of $v^4$ terms in the 
graviton scattering in Matrix theory compactified 
on 2-torus  
appeared in  \ci{bank}. Here we will ignore non-perturbative instanton 
contributions in $d=3$ SYM theory  which are  crucial for consistency 
of   type IIB string interpretation in the limit of vanishing 
volume of 2-torus \ci{bank}  but are not important in the present case.}
The corresponding  configuration in $D=10$  type II string  theory
is that of two parallel   $(2+0)$-branes 
(bound states of D2-branes and D0-branes). 
We  shall  find that the  $D=11$ supergravity  expression 
for the interaction potential 
between  two  membranes 
computed  using  the   procedure similar to that in 
  \ci{bbpt} (i.e. by  smearing the
 background produced by the source membrane  in the 
compact $x^-$  direction and fixing the  $p_-$ component of the momentum of the probe
membrane)
is in exact  agreement with the all-order  potential  
following from  the conjectured  Born-Infeld (BI) type  expression 
\ci{CT3} for  the leading large $N$, finite IR part of the quantum 
super Yang-Mills  effective action. 

At the same time, 
 the $D=10$ supergravity expression for the $(2+0)-(2+0)$ interaction 
potential is in correspondence  with the 
 SYM  expression  (and thus with the 
 light-like compactified $D=11$ expression) 
only  in  a  special   limit,  which turns out to be
 precisely the low-energy  limit  of \ci{maldacena}
($\a'\to 0$,  with  `Yang-Mills'  parameters 
fixed).  The result of taking this limit in the present  example  
is no longer 
equivalent simply to the   substitution $H \to H-1$ of  the harmonic function 
in 
the supergravity background as was the case 
 in the previously discussed 
`brane -- (bound state of branes)' interactions \ci{bbpt,CT3}.

It should be stressed that this test of supergravity-SYM correspondence 
is non-trivial: though it may seem that 
 the D2-probe action 
 has already a  BI form, part of the
 gauge field dependence is encoded in the 
curved space geometry produced by the source  D2-brane, 
and it is only after taking the limit that the resulting action becomes
 the  BI-type  expression  expected on the SYM side.

%%%%%%%%%%%%%%%%%%%%%%%%%%%%%%%%%%%%%%%%%%%%%
%%%%%%%%%%%%%%%%%%%%%%%%%%%%%%%%%%%%%%%%%%%%%%
%%%%%%%%%%%%%%%%%%%%%%%%%%%%%%%%%%%%%%%%%%%%%%%%

As we shall be using both the weak-coupling low-energy   IIA string theory
($D=10$ supergravity)  
picture and the light-like compactified M-theory ($D=11$ supergravity) 
 picture,  let us first review 
the relation between the corresponding  parameters 
%in the  Matrix theory (SYM)  context 
\ci{suss,bbpt,sen,seiberg}. 
The parameters  $\a'$, $g_s$  or $R_{11},$ $M_{11}$
\be
R_{11}=g_s (2\pi T)^{-1/2}\ , \ \ \ \ \  
 M_{11}= (2\pi g_s)^{-1/3} (2\pi T)^{1/2}
\ , \ \ \ \ \ \ 
R_{11} M^3_{11} = T\equiv { 1 \ov 2\pi \a'}    \ , 
\la{por}
\ee
of IIA theory 
 compactified on a circle $(x_{11}\equiv x_{11} + 2\pi R_{11}$) and a 
p-torus  of  volume $V_p$ 
and the parameters $R$,  $M_P$   of M-theory  with compact light-like  direction 
($x^- \equiv x^- + 2 \pi R$)  compactified on a p-torus  of  volume $\V_p$ are related as follows \ci{sen,seiberg}
\be
R_{11} M^2_{11} = R M^2_P\ , \ \ \  \ \ \  
V_p M_{11}^p = \V_p M_{P}^p \  , \ \ \ \ \ x M_{11}={\rm x} M_P  \ ,      
\la{para}
\ee
where $x$ and $\rm x$ are 
any  transverse scales   of the two theories. 
 The sector with $N$  D0-branes  in string theory 
or with momentum $p_-= { N \ov R}$ in M-theory 
is  described at low energies by the $U(N)$  SYM theory 
on the dual p-torus  with volume $\td V_p$
(dots stand for the standard adjoint scalar  and fermionic terms)
\be
 S =  -{ 1 \ov 4 \gym^2}  \int d^{p+1 } \td x\  \tr (F_{ab} F_{ab}) + ... \   , \ \ \ \ 
 \la{yyxx}
\ee
where $d^{p+1} \td x \equiv dt d^{p} \td x $ and
\be
 \gym^2 =  (2\pi)^{(p-1)/2}  T^{(3-p)/2}  \td g_s=
     (2\pi)^{-1/2}  T^{3/2}\td V_p   g_s
=  { (RM_P^2)^3 {\td V}_p }  \ ,       \la{coe} 
\ee
\be 
\td V_p={(2\pi)^p\ov T^p V_p} = 
{  (2\pi)^p \ov (R M_P^3)^{p} \V_p} \ ,  \ \ \ \ \   {\td V_p\ov \td g^2_s }
=  { V_p\ov  g^2_s}  \ .  
\la{deff}
\ee
The M-theory parameters $R, M_P, \V_p$ and  thus 
$\gym$ and $\td V_p$ 
remain finite in the limit $R_{11} \to 0, \ M_{11}\to \infty$  
or  $\a'\to 0, \ g_s \to 0, 
\ V_p\to 0$  \ci{seiberg,sen,maldacena}.

Consider two  parallel sets of D-branes 
(`probe' and  `source'  with charge $N$)
separated by a distance $r$ and imagine  computing  perturbative 
string theory diagrams
with one boundary on the probe brane and $L$ boundaries on the source brane.
%External legs may be open string states on branes or YM (collective %coordinates)  modes (F, X backgrounds).
  The large distance  limit of such
diagrams is expected to be  described  by  the 
massless closed string (supergravity) modes,
 while the  short distance  limit  -- by  the massless open string (SYM) modes. 
The contribution of the 
 $L=1$ (annulus) diagram  (in a  $F$=const background representing, e.g., 
 a  velocity or  a flux \ci{bachas}) 
expanded in powers of $F$ has, symbolically, 
 the structure ($n=7-p, \ \rr= {r\ov \a'}$)
$\ 
Z_1 \sim ~     {f^{(1)}_1\ov \rr^{n}} F^4  +    {f^{(1)}_2(\a'\rr^2)\ov \rr^{n+4}}F^6
 + ...\ $. 
As was  noted in \ci{dkps}, 
  $f^{(1)}_1$ has trivial (factor) dependence on $\a'$ (i.e. 
is independent of $\rr$)\foot{This is due to the fact that, as  explained in \ci{kiri},  only short open string  BPS multiplets  contribute
to the coefficient of $F^4$ term at one loop.}
and thus the leading $F^4/\rr^n$ term is the same for large and small $\rr$,  
so  that the $F^4$ 
terms  in  the supergravity and SYM expressions should agree. 
The result of \ci{bbpt} implies  that
the 2-loop string diagram  should give
$\  Z_2 \sim ~     {f^{(2)}_2\ov \rr^{2n}} F^6  +   
 {f^{(2)}_3(\a'\rr^2)\ov \rr^{2n+4}} F^8 + ...\ $, 
where $f^{(2)}_1=0$ and $f^{(2)}_2$ has trivial dependence 
on $\a'$. One may further conjecture \ci{CT3} that, in general, 
$\ Z_L\sim  ~    {f^{(L)}_{L} \ov \rr^{Ln} }  F^{2L+2} +   
{f^{(L)}_{L+1} (\a'\rr^2)\ov \rr^{Ln+4}} F^{2L+4} + ...\ $, 
where $f^{(L)}_{L} $ does not depend on $\rr$.
The $\a'\to 0$ limit of the open  string theory 
(with $F, \rr, \gym^2 \sim \a'^{(4-n)/2} \td g_s$ and modular UV cutoff being   fixed) 
 leads to   SYM theory,  so  that the  related conjecture 
about SYM theory is that 
the $F^{2L+2}$ terms appear in the large $N$,   IR part of the 
SYM effective action only 
at $L$-th loop order, i.e. that all  lower order 
$F^m$, $m < 2L+2$ terms at $L$-th loop 
have vanishing coefficients. 
The assumption that non-planar string 
diagrams (i.e. diagrams with closed string loops) are not included is justified
 provided the  string coupling  $\td g_s$ 
is small;  
the assumption that higher-order supergravity  $\a'$ corrections 
are not included is justified provided $N \td g_s \sim N \gym^2 $  is large, 
i.e. in the  large $N$ limit.
% (cf. \ci{maldacena}). 
To ignore subleading terms at each loop order one  needs 
to assume the low-energy limit, i.e. 
that $F^2/\rr^4 \ll 1$.
 To ensure that leading $F^{2L+2}/\rr^{Ln}$ terms are dominant, i.e. 
to be able to  ignore, say,  1-loop  SYM 
$F^6/\rr^{n+4}$ correction as compared to the 2-loop correction 
$N\gym^2 F^6/\rr^{2n}$  one is to assume that 
$N\gym^2/\rr^{n-4} \gg 1$ (which is the case if $N$ is large for fixed $\rr$).

Under  the above assumptions, 
%and  introducing  a constant scalar  background  with  scale ${\bf r}$, 
  the sum of the leading  large $N$ (planar)
   IR 
contributions to the quantum SYM effective action can be written as 
($F$ is  a gauge field background, $\rr$ is a scale of a 
scalar field   background) 
\ci{CT3,bbpt,ggr}
\be
\G   =\sum_{L=0}^\infty \G^{(L)} =  \ha \sum_{L=0}^\infty  \int d^{p+1} \td x 
  \  \bigg( {a_p  \ov  {\bf r}^{7-p}}\bigg)^L  
  (\gym^2 N)^{L-1} \  \hat  C_{2L+2} (F)  + ... \ ,  
\la{eff}
\ee
where we included also the tree-level $L=0$ term, 
\ $ a_p = 2^{2-p}\pi^{-(p+1)/2}
{\Gamma ({\textstyle{ 7-p \ov 2}})}$, 
\ $\hat  C_{2L+2} (F) \sim F^{2L+2}$ and dots stand for terms 
depending on covariant derivatives  and commutators of the gauge
field $F$ and scalars. 
It  was conjectured in \ci{CT3}
that 
$
 \hat  C_{2L+2} (F) =  
 \SSTr \ C_{2L+2} (F)$,  where $C_{2k}$ have the same Lorentz index structure as  the polynomials 
appearing in the expansion of the abelian BI action\foot{ 
Explicitly,  $C_0=1, \ C_2= - { 1\ov 4} F^2 ,$ \  $ 
C_4 = - { 1\ov 8}
 \big[F^4 - { 1\ov 4} (F^2)^2 \big], $ \ 
$ C_6 = - { 1\ov 12} \big[F^6 - { 3 \ov 8 } F^4 F^2  
+ { 1 \ov 32} ( F^2)^3\big] ,...,$ 
where $   F^k = F_{a_1a_2}
 F_{a_2a_3} ... F_{a_ka_1}.$}
and $\SSTr$ is a  modified symmetrized trace 
that reduces to the adjoint 
representation  trace for  some simple (abelian) backgrounds $F$.
 For $L=0,1$ the trace 
$\SSTr$ is  equal to the standard  symmetrized trace in  the adjoint
 representation; for $L=2$   its structure  was determined
using indirect considerations  in \ci{CT3}.

This  assumption is equivalent to the following conjecture
 for the derivative and commutator term  independent  part 
of the large $N$ effective  action  of maximally supersymmetric 
SYM  theory  \ci{CT3} (see also \ci{maldacena,kraus}) 
\be
\G= - { 1 \ov 2 N \gym^2} \int d^{p+1} \td x  \
 \SSTr \bigg( H^{-1}_p \bigg[\sqrt{-\det{\big(  \eta_{ab} I  +  H^{1/2}_p  F_{ab} }\big)}
- I \bigg] \bigg) \ ,  
\la{ohhh}
\ee
where
\be  H_p \equiv { a_p N \gym^2 \ov {\bf r}^{7-p}} \ .   
\la{hre}
\ee
This ansatz is consistent with  general   one-loop \ci{CT2,KT} and  some 
special two-loop \ci{bb,bbpt} perturbative 
calculations in SYM theory. 
Its correctness  is supported 
by the fact that this single expression
provides a  universal description of 
interaction potentials between  various  (bound states of) 
branes  computed using supergravity methods:
 (i) the $F^4$ term in \rf{ohhh}
gives the leading order ($1\ov r^{7-p}$)
 potentials for  BPS 
branes  with different amounts of supersymmetry 
(see, e.g., \ci{AB,lifmat,lif3,CT1,CT2})
as well as  for 
near-extremal branes \ci{mald1} and  non-supersymmetric  configurations of 
branes \ci{yank,tayl}; (ii) the $F^6$ term in \rf{ohhh}
gives 
subleading ($1\ov r^{2(7-p)}$) terms in the interaction potentials 
between brane configurations 
with 1/2, 1/4 and 1/8 fraction  of supersymmetry \ci{bbpt,CT3}.

Moreover, 
eq. \rf{ohhh}  reproduces the  exact ({\it all-order})   supergravity 
interaction potential  between two  0-branes  \ci{bbpt} (or two Dp-branes) 
   and a  0-brane and a non-marginal bound state
$(p+...+0)$ of D-branes \ci{CT2,CT3}, 
 in particular, the potential between a  0-brane and a (2+0)-brane or between 
a graviton and  a transversely boosted  membrane in $D=11$ \ci{CT3}.
  Other related arguments 
supporting the correctness of  the BI ansatz \rf{ohhh} were given in \ci{maldacena,dealwis,malda,ball}.
In particular, 
%\foot{Another argument for
 % the BI ansatz \rf{ohhh} can be given 
in the case of  $p=3$, 
the quantum  $N=4, D=4$ SYM effective action is expected  to 
have  special symmetry, reflecting the
fact that  the exact conformal invariance of this theory 
is spontaneously broken only by the 
adjoint scalar scale ${\bf r}$. Indeed, the abelian  version  of \rf{ohhh}
(its $\del X_m$  dependent  part) 
was shown to possess a kind of special conformal  symmetry 
\ci{malda,rena}.

Below we shall subject  \rf{ohhh}
 to a further  non-trivial test:  we shall show that it 
reproduces the  all-order  supergravity interaction
potential between two parallel $(2+0)$ branes in type IIA theory or 
two transversely 
boosted membranes  in M-theory  (with fixed values of $p_-$). 
This agreement is much less obvious than 
in the  $0 - (2+0)$  or graviton - membrane 
interaction case \ci{CT3}  and  depends on  details 
of the light-like  compactification prescription in $D=11$ 
or  details of the low energy limit in $D=10$.

In section 2 we shall find the 
explicit form of the interaction potential between two 
wrapped membranes in  Matrix theory  as implied by the  BI  
ansatz \rf{ohhh} for the SYM effective action. 
In section 3 we shall compare the SYM result  with the  all-order
 expression for the interaction potential in 
supergravity found using probe-source method. 
 We shall first   consider 
the  $(2+0)$--$(2+0)$  interaction potential  in $D=10$ supergravity
and show  its   agreement 
 with the SYM expression in the special  
 low energy limit of \ci{maldacena}. 
 We shall  then  compute  the interaction 
potential between two transversely boosted  wrapped 
membranes in $D=11$ supergravity with compact light-like direction
and demonstrate that performing the Legendre transformation 
$\dot x^- \to p_-$=fixed
(which,  in the present membrane context,  
 is a special case of the  $d=3$ world-volume 
   scalar-vector duality)  
 leads to the expression for the interaction
potential  coinciding with the  SYM  (BI)  expression.
The $D=11$  supergravity derivation is  more 
 straightforward than the $D=10$ supergravity one, 
 as it does not involve
  any special limit. This    illustrates  
 the 
%conceptual 
 advantage of 
 the  light-like compactification
procedure  of  \ci{suss,bbpt,seiberg}.  

As is well known, the  scalar-vector 
duality $x_{11} \to A_m$  transforms the standard $D=11$ membrane 
action \ci{bst} (in a $x_{11}$-independent $D=11$  supergravity  background) 
into the  D2-brane $d=3$  BI action  (in a  generic $D=10$ 
supergravity background) \ci{town,PP}. In Appendix  we  discuss 
the duality   transformation  of the  membrane action 
in the case when  it is  the {\it light-like coordinate} 
  $x^-$ that  is rotated  into a vector.

%%%%%%%%%%%%%%%%%%%%%%%%%%%%%%%%%%%%%%%%%%
\newsection{Membrane--membrane  potential from  super Yang-Mills theory}
%%%%%%%%%%%%%%%%%%%%%%%%%%%%%%%%%%%%
In  the   Matrix theory context, one is supposed to 
start with a  system  of  
0-branes in type IIA string theory 
 on a torus $V_2$  and  view  2-branes as  
 their `collective excitations'. 
Making $T$-duality which interchanges the numbers of D2-branes and D0-branes, 
 let us 
consider   
 the interaction of the two   $(2+0)$ bound states
wrapped over the dual  torus of volume $\td V_2$: 
one -- `probe'-- with    the $2$-brane number 
${\td n}_2=n_0$,  the $0$-brane number ${\td n}_0=n_2$ and the 
velocity in the direction $9$, and another 
-- `source' --  
with the $2$-brane number 
${\td N}_2=N_0$ and  the $0$-brane number ${\td N}_0=N_2$.
The   corresponding
pure gauge field    background   
  can be 
represented by the following 
gauge field strength matrices in the fundamental representation 
of $u(N )$\  ($N= n_0 + N_0)$
\be
\hat  F_{09} =  \pmatrix{   {\bf v} I_{n_0 \times n_0} & 0 \cr
  0  &    0_{N_0 \times N_0}\cr} \ , \ \ \ \  \ \ \ \ 
\hat  F_{12} =  \pmatrix{ {\bf f}_1 I_{n_0 \times n_0} & 0 \cr
  0  &   {\bf f}_2 I_{N_0 \times N_0}\cr} \ ,    
\la{tttr}
\ee
where the charges and the fluxes are related  by 
\be
2\pi n_2=n_0 {\td V}_2 \f_1 \ , \ \ \ \ \ \ 
  2\pi  N_2=N_0 {\td V}_2 \f_2 \ .  
\la{ttt} 
\ee
The background \rf{tttr} can be interpreted   as the
finite-$N$ Matrix theory configuration describing  the interaction of two 
$D=11$ membranes wrapped over the  torus of volume $\V_2$,  with the light-cone
momenta of the probe and the source membranes given by 
\be
p_{-}  ={n_0\ov R}=
{\T_2^{(1)} \V_2 \ov  \fff_1}={m_1\ov \fff_1}\ , \ \ \ \  
P_{-}  ={N_0\ov R} = {\T_2^{(2)} \V_2 \ov \fff_2}={m_2\ov \fff_2}\ .   
\la{defe}
\ee
Here the tensions are
\be 
\T_2^{(1)}={n_2 M_P^3\ov 2\pi}\ , ~~~~~\T_2^{(2)}={N_2 M_P^3\ov 2\pi}\ ,
\la{deef}
\ee
and the dimensionless fluxes $\fff_n$  and  the velocity of 
the probe membrane  are (see \rf{para}) 
\be 
 \fff_n  =  {\f_n \ov RM^3_P} \ , \ \ 
\ \ \ \ \ \ \      \v= {   {\bf v} \ov RM^3_P} \ .
\la{fluu}
\ee
 The relations of the type \rf{defe}, i.e. 
\be
p_{-}={\T_2 \V_2 \ov \fff}={m \ov  \fff}\ , \ \ \ \  \ \ \ 
m\equiv \T_2 \V_2 \ , \  
\la{FpR}
\ee
follow from the interpretation of the SYM (Matrix theory)  Hamiltonian as
the light-cone energy  \ci{bfss,suss} (see \rf{yyxx},\rf{coe}) 
\be
E_\tau={N{\td V}_2 {\bf f}^2 \ov 2(\gym^2)_{2+1} }={m^2\ov 2p_{-}} \ . 
\ee  
In the context of comparison 
with the $D=10$ string theory (supergravity)
 we shall have instead of \rf{fluu} the following fluxes and velocity 
\be
\ff_n = {\f_n \ov T} \ , \ \ \ \ \ \ \ v = {{\bf v}\ov T} \ ,    
\la{comp}
\ee
so that  the $D=10$ and $D=11$ parameters are related
as in \rf{para}\foot{Note that in view of \rf{por},\rf{para} 
the rescaling factor  in \rf{fluu} is equal to 
$ {RM_P^3}  = T { M_{P}\ov M_{11} }=T  ({R_{11} \ov R})^{1/2}$.}
\be
\ff_n = {M_P \ov M_{11}} \fff_n \ , \ \ \ \ \ \ \ v = {M_P \ov M_{11}}   {\v} \ .    
\la{compa}
\ee

It  is useful to subtract the traces and  describe  the background 
\rf{tttr}
by the  $su(N)$  matrices $F_{ab}$   which are proportional to the same 
 matrix $J_0$  as in \ci{CT3} 
\be
F_{09} = \f_{09} J_0=   {\bf v} J_0 \ ,  \ \ \ \  \ \ \ \ 
F_{12} =  \f_{12} J_0 = ({\bf f}_1-{\bf f}_2) J_0  \  ,   
\la{tra}
\ee
\be 
 J_0\equiv {1 \over n_0 + N_0  } 
 \pmatrix{  N_0  I_{n_0 \times n_0} & 0 \cr
  0  &   - n_0  I_{N_0 \times N_0}\cr} \  ,  \ \ \ \ { \tr } J_0=0 \ .  
\la{jjj}
\ee
 Since all of the components  of $\F_{ab}$  are proportional to
 the same matrix, 
the trace $\SSTr$ in \rf{ohhh} reduces simply to the trace in the adjoint representation.
Using that\foot{Given a diagonal matrix in the fundamental representation
of $u(N)$  with entries $a_i$ 
the corresponding matrix in the adjoint representation
has  entries $a_i-a_j$.
This  implies that $J_0$ has $ 2n_0 N_0 $ non-vanishing 
diagonal  elements
equal to $\pm 1$.}
 $\Tr J_0^{2k} =  2 n_0 N_0$, 
one finds that  polynomials constructed out of powers of $F_{ab}$
have the structure
 \be
\STr [ C_{2k}(F_{ab}) ]  =  \Tr [ C_{2k}(F_{ab}) ]  
 = 2 n_0 N_0 C_{2k}({\bf f}_{ab}) \ .
\la{rrry}
\ee
Substituting the background \rf{tra} into  $\Gamma$ \rf{ohhh} (and replacing
$N$ in \rf{ohhh}, \rf{hre} by $N_0$ to facilitate   comparison with  the 
supergravity probe-method  expression for the potential 
 which is linear in the probe's charge)
  we  get 
\be
\G= - { n_0  \ov   \gym^2} \int d^{3} \td x  \
   H^{-1}_2 \bigg[\sqrt{(1-H_2 {\bf v}^2)\big[1+ H_2 ({\bf f}_1-{\bf f}_2)^2\big]}
- 1  \bigg] \ ,   
\la{eeee}
\ee
\be  H_2 \equiv { 3 N_0 \gym^2 \ov 4\pi {\bf r}^{5}} \ .   
\la{vot}
\ee
The  SYM  scalar scale ${\bf r}$ will be   related to the scales 
$r$ and  r in the $D=10$ and $D=11$ supergravity expressions 
according to (cf. \rf{para},\rf{compa})
\be
r  = { {\bf r}\ov T}  \ , \ \ \ \ \ \  \r ={ {\bf r} \ov RM^3_P}\ , \ \ \ \ \ \ 
   r = { M_P \ov M_{11}} \r \ . 
\la{ses}
\ee
Below we shall reproduce the expression 
 \rf{eeee} as 
the action for a probe membrane  moving in a supergravity background 
of a  source membrane.  The   dependence of  the  interaction 
potential on the  {\it difference}  
of fluxes or  on the difference  of the  values of 
$p_-$ component of the momentum (cf. \rf{FpR}) 
  which is expected on the $D=11$ kinematics
grounds,   will  not be obvious a priori in the  exact 
supergravity expression derived   using the  probe-source method.
It will  appear  only after taking appropriate limit  in the $D=10$ 
supergravity expression for the $(2+0)-(2+0)$ interaction potential, 
or after a   duality (Legendre)  transformation  fixing  $p_-$ 
in the $D=11$ supergravity result for the  membrane 
action.

%%%%%%%%%%%%%%%%%%%%%%%%%%%%%%%%%%%%%%%%%
\newsection{Membrane   interaction  from  supergravity}
%%%%%%%%%%%%%%%%%%%%%%%%%%%%%%%%%%%%%%%%%%%%%%
Our starting point will be 
the $D=11$ supergravity background 
produced by a BPS membrane source \ci{dust}.
Applying a boost along $x_{11}$,
$x_{11}^{\prime}=x_{11}\cosh\beta -t\sinh\beta$,~~ $t^{\prime}=t\cosh\beta -x_{11}\sinh\beta$, \ 
we get
\be
ds_{11}^2=\K^{1/3}\big[\K^{-1}\big(-dt^{\prime 2}+dy_1^2+dy_2^2\big)+dx_{11}^{\prime 2}+dx_idx_i\big]\ ,   
\la{backk}
\ee
\be
~\K =1+\W \ , \ \ \ \ \ \ \W= {\Q\ov r^6}\ ,~~~~
\Q={8 N_2\ov M_P^6} \ ,   
\la{ddd}
\ee
\be C_{t^{\prime} y_1y_2}=\K^{-1}-1\ ,~~~~~~
C_{t y_1y_2}=(\K^{-1}-1)\cosh\beta ,~~~ 
C_{11 y_1y_2}=-(\K^{-1}-1)\sinh\beta \ .   
\la{ccc}
\ee
The  components of the metric  
in terms of the light-cone coordinates 
$x^{\pm}=x_{11}\pm t$ or $ \tau = \ha x^+$ and $x^-$ are
$$
g_{\tau\tau}={e^{-2\beta}}\K^{-2/3}(\K-1)\ ,~~~~g_{--}={\textstyle {1 \ov 4}} {e^{2\beta}}\K^{-2/3}(\K-1)\ ,
$$
\be
g_{\tau -}= \ha (1+\K)\K^{-2/3},~~~~~~g_{y_1y_1}=g_{y_2y_2}=\K^{-2/3}\ , 
~~~~~~~g_{ij}=\delta_{ij} \K^{1/3}. 
\la{meta}
\ee
We shall consider  either space-like 
($x_{11} \equiv  x_{11}+2\pi  R_{11}$)
or light-like ($x^{-}\equiv  x^{-}+2\pi  R$)
compactification  and  smear the membrane  background 
in the compact  `transverse'  direction
(this corresponds  to fixing the component of the 
11-dimensional  momentum of the source membrane). Since the above membrane 
solution is 
a BPS one, this is equivalent  to `smearing' the harmonic function $\K$
in the compact direction.
In the case of the  
space-like compactification  we get
$$
\W \to \Q_{11}\sum_{n=-\infty}^{\infty} (x_{11}^{\prime 2}+r^2)^{-3} =
\Q_{11} \sum_{n=-\infty}^{\infty} \bigg( [(x_{11}+2\pi n R_{11})\cosh\beta -t\sinh\beta 
]^2+r^2\bigg) ^{-3}
$$
\be
\rightarrow \ \ \W_{11}=  {\Q_{11} \ov 2\pi  R_{11} \cosh\beta}\int_{-\infty}^{\infty} {dx_{11}\ov (r^2+x_{11}^2)^3} =
{3\Q_{11}  \ov 16 R_{11} \cosh\beta}{1\ov r^5}
\ , 
\ee
\be
\K_{11} =1+\W_{11} =1+{3\Q_{11} \ov 16 R_{11}\cosh\beta \ r^5} \ ,  \ \ 
\ \ \   \Q_{11} = { 8 N_2 \ov M^6_{11}}   \ .  
\la{Wone}
\ee
In the case of 
compactification on the light-like  direction $x^{-}$
$$
\W \to  \Q\sum_{n=-\infty}^{\infty} 
(x_{11}^{\prime 2}+\r ^2)^{-3}=
\Q\sum_{n=-\infty}^{\infty} 
\bigg(\big[ \ha e^{\beta}(x^{-}+2\pi n R)+
\ha e^{-\beta}x^{+} \big]^2+\r ^2\bigg)^{-3} 
$$
\be 
\rightarrow\ \  \W_-=  {\Q\ov \pi  R e^{\beta}}
\int_{-\infty}^{\infty} {dx^{-}\ov [\r ^2+(x^-)^{2}]^3} =
{3\Q \ov 8 R e^{\beta}}{1\ov \r ^5}\ , 
\ee
\be
\K_{-}=1+\W_{-}=1+{3 \Q  e^{-\beta}\ov 8R \r ^5} \ .    
\la{mmm}
\ee
%%%%%%%%%%%%%%%%%%%%%%%%%%%%%%%%%%%%%%%%%%%%%%
\subsection{$(2+0)$--$(2+0)$ interaction in   $D=10$
  supergravity}
%%%%%%%%%%%%%%%%%%%%%%%%%%%%%%%%%%%%%%%%%%%%%%%%%%%
The  $D=10$ type IIA supergravity background \ci{RT}
representing the bound state
 $(2+0)$ 
of $N_0$  D0-branes  and  $N_2$ D2-branes  wrapped over 
 a torus of volume  $V_2$     can be  obtained, e.g., 
by compactifying the boosted M2-brane background along the spatial direction 
$x_{11}$    \rf{backk},\rf{Wone}
%\foot{We  use string-frame metric and
%change  the sign of $B_{mn}$ compared to \ci{RT}, 
% $B_{y_1y_2} =- C_{11 y_1y_2}$. } 
 $$
ds^2_{10} =   K^{1/2} [ - K\inv dt^2 + 
\Kt\inv (dy_1^2 + dy_2^2)    +  dx_i dx_i ]  \ ,  $$
$$ e^{2\phi }=   K^{3/2}  \Kt\inv   \ , \ \  \ \ \ 
C_{t y_1y_2} = - \sin \t \ W \Kt\inv \ , $$
$$ C_{t }  =  - \cos \t\ W K\inv \ , \ \  \ \ \ 
  B_{y_1 y_2 }  = \sin \t\  \cos \t\  W \Kt^{-1}  \ .   $$
\be  K=1+W\ ,~~~~~~~{\Kt}=1+W\sin^2\theta=\K_{11}
\ ,     \la{boo}
\ee
where
\be
W={Q^{(2)}_0\ov r^5} \sqrt{1+  f_2^2},~~~
\cos\theta ={ 1 \ov \sqrt{1+  f_2^2}},~~~~\sin\theta ={ f_2
 \ov \sqrt{1+  f_2^2}}\ , 
\la{Wq}
\ee
$$
  f_2={ Q_2\ov  Q_0^{(2)}} = { V_2 T  N_2 \ov  2\pi  N_0}=
{ 2\pi   N_2 \ov   \td V_2  T  N_0}\ ,   
 \ \ \  \   
Q_2={3N_2  g_s\ov 2(2\pi)^{1/2}T^{5/2}}   ,~~~~~Q_0^{(2)}=
{3(2\pi)^{1/2} N_0 g_s\ov 2 T^{7/2} V_2} . 
$$
The angle $\theta$ is related to the 11-dimensional  boost parameter
$\beta$ by 
$\sin \t = (\cosh \beta)^{-1}$.
The limit of  $\sin \t=0$ ($\Kt=1$)  or  $f_2\to 0 $ ($N_0 \gg N_2$) 
corresponds to the 0-brane background smeared over the volume $V_2
$  ($Q_0^{(2)}$ is the  effective charge parameter 
of 0-brane background). 
The limit of 
 $\sin \t =1$ ($K=\Kt$) or $f_2 \to \infty$  ($N_2\gg N_0$) 
 corresponds to  the  pure 2-brane background. 

Having in mind comparison with  Matrix theory, 
we have presented  the $(2+0)$  background from the `0-brane point of view',
 i.e.  as a modification (due to the presence of a D2-brane charge)
of the D0-background smeared over the torus $V_2$.
To establish  the  correspondence   with the  SYM theory 
 one is then to   consider the T-dual 
theory. $T$-duality along the two directions of the torus
 transforms  the original  
theory  with coupling $g_s$  and 
  $(0+2)$ ($N_0,N_2)$  bound state wrapped over  $V_2$ into the 
 the  dual theory with coupling $\td g_s$ 
($V_2/g^2_s= \td V_2/\td g_s^2$)
and the 
  $(2+0)$  ($\td N_2=N_0, \ \td N_0=N_2$) 
bound state  wrapped over the dual torus with volume 
 $\td V_2 =(2\pi/T)^2 V_2\inv$.

Applying $T$-duality along  $(y_1,y_2)$ 
one finds that 
  the  $T$-dual background  has  (apart from the change of sign of $B_{mn}$) 
exactly  the same form as  \rf{boo}
 but 
with  
$\sin \t \leftrightarrow \cos \t$, i.e., in particular,  with\foot{The transformation rules in \ci{dual}
imply that $\td C_t = C_{t y_1y_2} - C_t B_{y_1y_2}$, etc.} 
\be 
\Kt \to \td 
\Kt=1 + W \cos^2\t \ .    
\la{KKK} \ee
 As one might  expect, this  transformation  is equivalent to replacing
 $f_2 $
by $\td f_2 = 1/f_2$ or $N_0 \leftrightarrow N_2$, 
$V_2 \to \td V_2$ 
  (as well as changing $Q^{(2)}_0\to Q_2 $ in $W$  \rf{Wq} 
as   $W$  is to remain  invariant).  

The 
$T$-duality transforms also the  $(0+2)$ ($n_0,n_2)$ probe  wrapped over $V_2$ in the 
original theory    into the $(2+0)$ ($\td n_2=n_0, \ \td n_0=n_2)$
probe  wrapped over $\td V_2$ in the dual theory. 
The action of a   D2-brane  probe 
 propagating in  the dual  type IIA supergravity background 
   is   (we use the static gauge;  
$m,n=0,1,2; \ i,j=3,...,9$)
 $$
 \td I_2 = -  \td T_2 \int d^3  \td x  \bigg[  e^{-\td \phi} \sqrt {-\det\big( \td G_{mn } +
 \td G_{ij} \del_m x^i \del_n x^j +  \td \FFF_{mn} \big)}
$$
\be 
    - \ {1\over 6} e^{mnk}\td  C_{mnk}  -  {1\over 2}  \e^{mnk}\td  C_m\td  \FFF_{nk}  \bigg] \ ,   
\label{act}      
    \ee
where in  the present context of $T$-dual theory 
$$
\td  \FFF_{mn} \equiv  T^{-1}  \td F_{mn} + \td B_{mn}  \ ,   \ \ \ \ \
\td T_2 = \td n_2 \td g_s^{-1}(2\pi)^{-1/2}T^{3/2}= {n_0 (2\pi T)^{1/2}\ov g_s \td V_2} \ . $$
    To find  the $(2+0)$ probe action one should introduce  a constant magnetic field $\td F_{12} $  proportional to $\td n_0 = n_2$ (for  a similar  discussion of $0-(2+0)$ interaction see  \ci{CT1,CT3}).

 Substituting the   $(2+0)$ source background  ($T$-dual of \rf{boo}) 
into the action of the $(2+0)$ probe 
we get (we  ignore the dependence on the spatial coordinates parallel to the brane) 
$$
\td I_2 = - \td  T_2  \int d^3 \td x  \bigg[(K \tilde \Kt )^{-1/2} \sqrt{(1-K v^2)(1+K^{-1} \tilde \Kt^2\td {\cal F}^2 ) } $$
\be
+  \ W  {\tilde \Kt}^{-1} \cos \theta + W {\td {\cal F}}  K^{-1}\sin \theta \bigg]
\ ,   
\la{aca}
\ee
where 
\be
\td{\cal F}=\td \FFF_{12} = {  f_1}  -W {\tilde \Kt}^{-1}\sin \theta \cos\theta\ ,~~~~~~~
   f_1 ={T\inv \td  F_{12} } = { 2 \pi \td n_0 \ov \td V_2 T \td n_2}
 \ , \ \ \ \  v= \del_t x_9 \ .  
\ee
The system of parallel D2-branes wrapped over $\td V_2$ at low energies 
 is expected to be described  by 
the  SYM theory on $\td V_2$ \ci{witt}. 
Let us show that  in the low-energy or   `Yang-Mills' 
limit of \ci{maldacena}
  this complicated-looking supergravity action  indeed  reduces 
 to  effective action \rf{eeee}
of SYM theory     on   $\td V_2$ 
with the SYM coupling given in  \rf{coe}. 
Expressing the parameters in the action \rf{aca} 
 in terms of the SYM  parameters 
$\f_n, \ {\bf v}, \ {\bf r}$ and $ \gym$ 
(see \rf{comp},\rf{ses})
\be
f_n = T^{-1} \f_n \ , \ \ \ \ \ v = T^{-1} {\bf v} \ ,   \ \ \ \  r= T\inv {\bf r} \ ,   \ \ \ \  
 \gym^2 = (2\pi)^{1/2}  T^{1/2}  \td g_s = { \td n_2 T^2 \ov \td T_2} 
\ , 
\la{compe}
\ee
and taking the low-energy (or short-distance) limit \ci{maldacena} \ 
 $T\to \infty$ (or $\a'\to 0$)
with $\f_n, \ {\bf v}, \ {\bf r}, \ \gym$ being fixed,  we find   
that $\sin\t =1 + O(T^{-2})  ,  \ 
\cos\t =T^{-1} \f_2  +   O(T^{-3})$ and 
\be 
K = 1 + H_2 T^2 \to H_2 T^2  , \ \ \ 
\tilde \Kt = 1+ H_2 T^2  (1 + O(T^{-2})) \to H_2 T^2  ,  \ee   $$
\td {\cal F}= T^{-1} ( \f_1 - \f_2) +   O(T^{-3}) \ ,   \ \ \ \ \ \ \ \
H_2 \equiv   { 3 N_0 \gym^2 \ov 4 \pi  {\bf r}^{5}}  \ . $$
Expanding \rf{aca} in  powers of inverse string tension $T\inv$, we 
 finish with 
$$
\td I_2 =-  { \td T_2 \ov T^2} \int d^3 \td x 
  \
\bigg[   H^{-1}_2 \sqrt{(1-H_2 {\bf v}^2)\big[1+ H_2 ({\bf f}_1-{\bf f}_2)^2\big]}
$$
\be
  +  \  T^2 + \f_1 \f_2 - \ha \f_2^2  -  H^{-1}_2     + O( T^{-2})   \bigg] \  
\la{yes}
\ee
$$
= - \td  T_2 \int d^3\td  x +  {\td  T_2 \ov T^2} \int d^3\td  x\   \bigg[ \ha v^2  - \ha \f^2_1 - {\cal V}  +O( T^{-2})  \bigg]   \  , 
\ \ \ \ \  { \cal V } = O({1 \ov  {\bf r}^{5} }) \ . 
$$
Since  $ { \td T_2 \ov T^2} ={  n_0 \ov \gym^2}$  (see  \rf{compe}),  the finite part of  
 this  action is
equivalent to the SYM  BI-type expression  
\rf{eeee} (up to the  constant  `self-energy' $O(\f^2)$ term).

We would like to stress that this  result  is 
non-trivial:

 (i) though it may seem that  the probe action \rf{act}
 has already   a  BI-type form, it  also  contains a complicated dependence
on the source flux parameter in the background supergravity fields:
the gauge field enters  not only  through $\td {F}$
but also through, e.g., the antisymmetric tensor field  in $\td {\cal F}$, 
i.e. the gauge field  background  is partially encoded in the geometry; \ 

  (ii) it is only 
   in  the   special low energy  limit ($T\to \infty$)   that 
 the two fluxes combine  to form the difference  appearing in  the 
SYM-BI action 
\rf{eeee}.

This all-order agreement between the supergravity and SYM results 
for the $(2+0)--(2+0)$  interaction potential provides a
 check of the consistency  of the low energy 
SYM limit of \ci{maldacena} and 
of the BI ansatz \rf{ohhh}.

%implicitly (in the context of the discussion of the next section) also 
%of the limit of \ci{seiberg}.

As expected, a  similar  agreement if found also for 
 the $(2+0)-(2+0)$  system in the `pre- T-duality' picture, 
i.e. when one considers  the 
two D2-branes and describes their D0-brane content  by  the 
magnetic fluxes directly related  to the SYM fluxes.
In this case the $2+0$ background is described by \rf{boo}
with $f_2 \to 1/f_2$ (i.e. with $W={Q_2\ov r^5} \sqrt{1+  f_2^2}, 
\ \sin\theta ={ 1
 \ov \sqrt{1+  f_2^2}})$.
Then $f_2$ and $f_1$ in the  corresponding probe action 
(\rf{aca}  with $\sin \theta \leftrightarrow \cos \theta$, $T_2 = 
   n_2 g_s^{-1}(2\pi)^{-1/2}T^{3/2} $) 
are  proportional to the SYM field strength, 
so that the above  limit   gives again \rf{yes}.

%%%%%%%%%%%%%%%%%%%%%%%%%%%%%%%%%%%%%%%%
\subsection{Membrane -- membrane interaction in 
$D=11$  supergravity with compact  light-like direction}
%%%%%%%%%%%%%%%%%%%%%%%%%%%%%%%%%%%%%
We  shall now  demonstrate that  the  equivalence 
between the SYM  and supergravity results  for the membrane--membrane potential
can be established  directly 
(with no need to take the special limit of the supergravity expression) 
in the framework
of the  $D=11$ supergravity with compact light-like  direction, 
in agreement with previous suggestions 
  \ci{suss,bbpt,seiberg}.

 The action  of a M2-brane  probe with tension $\T_2^{(1)}$ \rf{deef} 
propagating in  curved  $D=11$ background  is  
%\ci{memact}
(we use the static gauge with $\tau= \ha x^+$) 
\be
S_2=-\T_2^{(1)} \int d^3x \bigg[\sqrt{-\det (g_{mn} +
\partial_{m}x^i\partial_{n}x^j g_{ij} )}
-C_{\tau  x_1 x_2 }\bigg]\ .
\la{ppp}
\ee
In the case of the background produced by 
a boosted membrane source  averaged in $x^-$ direction
 we get (see \rf{meta},\rf{ccc},\rf{mmm})
\be
C_{\tau x_1x_2 } = C_{tx_1x_2} { dt \ov d \tau} +  C_{11 x_1x_2} { dx_{11} \ov d \tau}
=
(e^{-\beta}-\ha {{\dot x^{-}}}e^{\beta})(\K^{-1}_- -1) \ ,  \ \ \ \ \ 
\dot x^{-} = \del_\tau x^- \ , 
\la{cccc}
\ee
and thus ($\v_i = \del_\tau x_i$, \ $\K_- = 1 + \W_-$)
\bea
S_2&=&-\T_2^{(1)} \int d^3x \bigg[\K^{-1}_-\sqrt{-[e^{-2\beta}\W_- +(2+\W_-){\dot x^{-}}+{\textstyle{1\ov 4}}e^{2\beta}\W_- ({\dot x^{-}})^2+\K_-\v^2 ]}\non
&&-\ (e^{-\beta}-{\ha {\dot x^{-}}}e^{\beta})(\K^{-1}_--1) \bigg] \ . 
\label{action}    
\eea
Since we  are going  to  consider the probe with 
fixed value of the light-like momentum $p_-$,  we are to perform, as in \ci{bbpt},
 the Legendre transformation 
$\dot x^- \to p_-$ and set 
$p_-$ to be constant. From a 
more  general point of view, this transformation is a   special case 
of a $d=3$ world-volume  duality transformation 
that rotates a scalar into a vector (and, in the case of 
  space-like $x_{11}$-compactification, 
relates the  M2-brane action to  the  D2-brane action  \ci{town}). 
The transformation $x^- \to A_m$  is  discussed in Appendix, 
where it is demonstrated  that  $p_-$ has the interpretation  of the  
(inverse of)  
magnetic field strength. 

Namely,  let us 
assume that the membrane coordinates depend only on $\tau$
and 
compute $S'_2 = \int d \tau  L',$ \   $  L'= L (x^-(p_-)) -    \dot x^- p_-$.
As a result,   
$$
L^{\prime}={4e^{-2\beta}p_{-}\ov \W_- }\bigg(1-\sqrt{(1-{\v^2 \W_- \ov 4e^{-2\beta}})[1+{(2e^{-\beta}p_{-}- m_1)^2\W_- \ov 4e^{-2\beta}p_{-}^2}}]\bigg)
$$
\be +\ 2e^{-\beta}(e^{-\beta}p_{-}- m_1) \ , \  \ \ \ \ \  m_1 = \T_2^{(1)}\V_2 \ .   
\la{accc}
\ee
Let us introduce (see \rf{mmm},\rf{ddd})
\be
{\rm H}_2={\W_{-} \ov 4e^{-2\beta}}={\W_{-} \ov \fff_2^2}={3\Q N_0\ov 16 R^2 \T_2^{(2)}\V_2 \r^5}={3 \pi N_0\ov R^2 M_P^9 \V_2 \r^5}\ 
,  
\label{hF}   
\ee
and 
\be \fff_2 = 2 e^{-\beta}= { m_2 \ov P_-} \ .     
\la{Fp}
\ee
$N_0$ is related to the boost (with parameter  $\beta$) 
 applied to the source membrane
and $P_-$ is the  momentum of the source.
Both membranes are assumed to be wrapped over the torus 
with volume $\V_2$.
The relation \rf{Fp} between the dimensionless 
flux $\fff_2$  corresponding to the source membrane 
and the boost $\beta$  can be understood as follows: 
with the  choice of the 
time variable $\tau=\ha x^{+}$, the  light-cone  Hamiltonian and momentum are 
\be
E_{\tau}= E-P_{11}=m e^{-\beta}\ ,~~~~~~~~~~ P_{-}=\ha (E+P_{11})
=\ha m e^{\beta} \ ,  
\la{some}
\ee
so that (cf. \rf{FpR})
$\fff={m\ov P_{-}} =2 e^{-\beta}$.

Using \rf{hF} and \rf{Fp}  and introducing 
the flux corresponding to the probe membrane
\be
\fff_1=  { m_1 \ov p_-} \ ,   
\la{ffff}
\ee
we can rewrite the probe Lagrangian  \rf{accc}  in terms of 
$\fff_1$ and $\fff_2$  
$$
L'
= {m_1 \ov \fff_1 } \bigg[{\rm H}_2^{-1}\bigg(1-\sqrt{(1-{\rm H}_2 \v^2)
[1+{\rm H}_2(\fff_1-\fff_2)^2 ]}\bigg) 
    + \ha  \fff^2_2 - \fff_1 \fff_2 \bigg]  $$
\be
=\   p_- ( \ha \v^2 - \ha \fff_1^2 - {\cal V})=
 - { m_1^2 \ov 2 p_-}   + {  p_-  \v^2 \ov 2}  - p_- {\cal V}  \ , 
\ \ \ \ {\cal V} =  O({1 \ov {\rm r}^5}) \ .
\la{acty}
\ee
The constant terms here are in agreement with the  general 
dual ($x^- \to A_m$)  form  of the membrane action  found in Appendix.

Using, finally,  the relations \rf{fluu},\rf{ses},\rf{vot}
we observe  that 
\be
{\rm H}_2 \v^2=H_2 {\bf v}^2,~~~~~~{\rm H}_2 (\fff_1-\fff_2)^2=H_2
 ({\bf f}_1 -{\bf f}_2)^2,~~~~~~
{m_1\ov \fff_1 {\rm H}_2}={n_0 \td V_2\ov \gym^2 H_2} \ , 
\ee
and thus conclude  that 
 \rf{acty} is equivalent to  the $D=10$ action \rf{yes} as well as to 
the SYM  expression for the potential   in \rf{eeee}.

%%%%%%%%%%%%%%%%%%%%%%%%%%%%%%%%%%%%%%%%
%%%%%%%%%%%%%%%%%%%%%%%%%%%%%%%%%%%%%%%%%%%%%%%%%%%
%%%%%%%%%%%%%%%%%%%%%%%%%%%%%%%%%%%%%%%%%%%%%%%
\bigskip
\centerline {\ \bf Acknowledgments}
%\end{center}
The work of I.C. was supported in part  by NSF  grant PHY-9309888. 
A.A.T.  would like to thank A.M. Polyakov for useful comments and 
acknowledges  also the support
 of PPARC and  the European
Commission TMR programme grant ERBFMRX-CT96-0045.
%%%%%%%%%%%%%%%%%%%%%%%%%%%%%%%%%%%%%%%%%%%%%%%%%%%%%%%%%%%

\bigskip

\setcounter{section}{0}
\setcounter{subsection}{0}

\appendix{Light-like scalar -- vector duality
 % Duality 
transformation  
%\ $x^- \to A_m$ \ 
of  
%the $D=11$ 
membrane action}
%%%%%%%%%%%%%%%%%%%%%%%%%%%%%%%%%%%%%%%%%%%%%%%%%%%%%%%%%%%%
\setcounter{equation}{0}

%%%%%%%%%%%%%%%%%%%%%%%%%%%%%%%%%%%%%%%
The $d=3$ duality transformation $x_{11} \to A_m$ is known 
to relate the flat-space 
 Nambu-type  membrane action
 $\int \sqrt{-\det_3\ (\del_m x^M \del_n x^M) }$ and the  Born-Infeld
   D-membrane action\ \ \ \ \ \ \ \ $ \int \sqrt{-\det_3\
 (\del_m x^\a \del_n x^\a + {{\rm  F}}_{mn})  }$ \ci{town}. 
If one considers instead the duality transformation $x^- \to A_m$
the result is quite different  as we  find below. 
 This  new dual action 
can probably be viewed as a special singular limit 
of the curved-space D2-brane BI action.
 We  suspect  it  may 
 have some 
interesting applications, apart from being 
the free part of the action \rf{acty} 
derived in section 3.2.

Let us start with   the membrane action  in flat  $D=11$ 
background ($ds^2_{11} = dx^+dx^- + dx_i dx_i$),  
choosing  the static gauge with $\tau = \ha x^+$ ($m,n=(\tau, a), \ a,b=1,2$)
\be
S_2=-\T_2  \int d^3x \ \sqrt{-\det_3\   h_{mn} }\  \to \  
\ha \T_2  \int d^3x  \bigg[   U  h_{\tau\tau} \det_2 ( h_{ab} -  h_{\tau\tau}\inv 
h_{\tau a}h_{\tau b})    -  U\inv \bigg] 
\ . \ 
\ee
Here $U$ is an auxiliary field introduced  to  `linearise' 
the square root  and 
$$
h_{\tau\tau}  =  2 \del_\tau x^-  + \del_\tau x^i \del_\tau x^i \ , 
\ \ \ 
h_{\tau a}  = \del_a x^- +  \del_\tau x^i \del_a x^i \ , \ \ \ 
h_{ab}  = \delta_{ab}  +  \del_a x^i \del_b x^i \  . 
$$ 
To perform the duality transformation $x^- \to A_m$ 
we are to replace $\del_m x^-$ by a vector $\L_m$,  add
 the Lagrange multiplier term
\be
-\ha \T_2 \int d^3 x \ \ep^{mnk} \L_m {{\rm  F}}_{nk}
= -\T_2 \int d^3 x \ ( \L_\tau {{\rm  F}}  + \L_a {{\rm  F}}^a )  \ , 
\ee  $$
{{\rm  F}}_{nk} = \del_n A_k - \del_k A_n \ ,  \ \ \
{{\rm  F}}= {{\rm  F}}_{12} \ , \ \ {{\rm  F}}^a = \ep^{ab} {{\rm  F}}_{b\tau} \ , 
$$
and   `intergate out'  $\L_m$. 
Since the induced metric $h_{mn}$ depends on  $ \del_m x^-$ 
only linearly,  
it is useful to redefine $\L_m\to \l_m$\   ($\L_\tau  =  \ha \l_\tau - \ha \del_\tau x^i \del_\tau x^i, 
\  \L_a = \l_ a  - \del_\tau x^i \del_a x^i$)
 so that $h_{\tau\tau}  = \l_\tau, \
h_{\tau a}  = \l_a$. Then 
\be
S'_2=  \ha \T_2  \int d^3x  \bigg[ U \l_\tau  \det_2 ( h_{ab}
 -   \l_{\tau}\inv  \l_{ a}\l_{ b})    -    U\inv 
-    ( \l_\tau  -  \del_\tau x^i \del_\tau x^i) 
 {{\rm  F}}  -  2  ( \l_ a  - \del_\tau x^i \del_a x^i)  {{\rm  F}}^a \bigg] 
\ . 
\ee
Since $$
 \det_2 ( h_{ab}-  \l_{\tau}\inv  \l_{ a}\l_{ b}) = h 
(1 - \l_{\tau}\inv  h^{ab} \l_{ a}\l_{ b}) \ , \ \ \ h\equiv  \det_2\ h_{ab} \ ,
 $$
it is now easy to integrate out $\l_a$, 
\be
S'_2=  \ha \T_2 \int d^3x  \bigg[   U h \l_\tau   -
 {{\rm  F}}  \l_\tau     -   U\inv 
  + U\inv  h\inv   h_{ab} {{\rm  F}}^a {{\rm  F}}^b 
   +  \del_\tau x^i \del_\tau x^i {{\rm  F}}     +   2 \del_\tau x^i \del_a x^i {{\rm  F}}^a \bigg] 
\ . 
\ee
Solving for  $\l_\tau$  gives  $ U   h =  {{\rm  F}}$;
eliminating $U$,  we finally obtain
\be
S'_2=  \ha \T_2 \int d^3x  \bigg[ -  {{\rm  F}}\inv  \det_2 (\delta_{ab}  +  \del_a x^i \del_b x^i) 
 +  {{\rm  F}}  \del_\tau x^i \del_\tau x^i
     +   2 {{\rm  F}}^a  \del_\tau x^i \del_a x^i 
\ee
$$
 +  \  {{\rm  F}}\inv   {{\rm  F}}^a {{\rm  F}}^b (\delta_{ab}  +  \del_a x^i \del_b x^i)\bigg] \  , 
$$
or, equivalently, 
\be S_2' 
=  \ha  \T_2 \int d^3x  \bigg[ -  {{\rm  F}}\inv   \det_2 (\delta_{ab}  + \del_a x^i \del_b x^i)  
+    {{\rm  F}}\inv   {{\rm  F}}^a {{\rm  F}}_a  + 
 {{\rm  F}}\inv ( {{\rm  F}}^a \del_a x^i + {{\rm  F}}  \del_\tau x^i)^2 \bigg] \ . 
\la{aaaa}
\ee
The last term can be rewritten also as $(\ha\ep^{mnk} {{\rm  F}}_{mn} \del_k x^i)^2$. 
When $\del_a x^i=0$ and ${{\rm  F}}_{mn}$=const  this action 
 gives the first two terms in the action  
\rf{acty} ($S'_2= \int d\tau L_2'$) 
with $p_- ={\T_2 \V_2 {{\rm  F}} }$. The magnetic field 
 ${{\rm  F}}$  is thus the  inverse of the 
flux  $\rm f$ \rf{ffff}, which is proportional to the SYM flux \rf{fluu}.

%%%%%%%%%%%%%%%%%%%%%%%%%%%%%%%%%%%%%%%%%%%%%%%%%%%%%%%%%%%%%%%%%%%%%%5


\begin{thebibliography}{9}

\bibitem{dkps}
M.R. Douglas, D. Kabat, P. Pouliot and S.H. Shenker,
 \np B485 (1997) 85,
hep-th/9608024.



\bibitem{bfss}
 T. Banks, W. Fischler, S.H. Shenker and L. Susskind,
\pr D55 (1997) 5112, hep-th/9610043.


\bibitem{AB}
O. Aharony and M. Berkooz,
{\it Membrane Dynamics in M(atrix) Theory}, 
 Nucl. Phys. B491 (1997) 184,
hep-th/9611215.

\bibitem{lifmat}
G. Lifschytz and S.D. Mathur,
{\it Supersymmetry and Membrane Interactions in M(atrix) Theory},
\np B507 (1997) 621, 
hep-th/9612087.

\bibitem{lif3}
G. Lifschytz,
{\it Four-Brane and Six-Brane
 Interactions in M(atrix) Theory},
hep-th/ 9612223.


\bibitem{CT1}
I.~Chepelev and  A.A.~Tseytlin,
{\it Long-distance  interactions of D-brane  bound states
and longitudinal 5-brane in M(atrix) theory },
Phys. Rev. D56 (1997) 3672, 
hep-th/9704127.

\bibitem{CT2}
I.~Chepelev and  A.A.~Tseytlin,
{\it Interactions of type IIB D-branes from D-instanton matrix model},
hep-th/9705120.

\bi{KT}
D.~Kabat and W.~Taylor,
{\it Linearized supergravity from Matrix theory},
 hep-th/9712185.

\bibitem{bbpt} K.~Becker, M.~Becker, J.~Polchinski, and A.A.~Tseytlin, 
{\em Higher order graviton scattering in M(atrix) theory}, 
Phys. Rev. D56 (1997) 3174, 
hep-th/9706072.

\bi{suss}
L. Susskind, {\sl Another Conjecture about 
M(atrix)
Theory}, hep-th/9704080.


\bibitem{CT3}
I.~Chepelev and  A.A.~Tseytlin,
{\it  Long-distance interactions of branes: correspondence  between 
supergravity  and super Yang-Mills descriptions},
hep-th/9709087.


%\bi{T} A.A. Tseytlin, ``Interactions between branes and matrix theories", 
%Contribution to Strings '97, hep-th/9709123.


\bi{dps}
M. Douglas, J. Polchinski and A. Strominger,
{\it Probing Five-Dimensional Black Holes with D-Branes}, 
hep-th/9703031.

\bi{maldacena}
J. Maldacena, {\it Branes probing Black Holes}, Contribution
to Strings '97, hep-th/9709099.


\bi{sen}
A.~Sen,
{\it $D0$ Branes on $T^n$ and Matrix Theory},
hep-th/9709220.
 
\bi{seiberg}
N.~Seiberg,
{\it Why is the Matrix Model Correct?},
Phys. Rev. Lett. 79 (1997) 3577,
hep-th/9710009.
 
\bi{hell}
S. Hellerman and J. Polchinski, {\it  Compactification 
in the light-like limit}, 
hep-th/9711037. 

\bi{PP}  
J.~Polchinski and P.~Pouliout,
{\it Membrane Scattering with M-Momentum Transfer},
Phys. Rev. D56 (1997) 6601,
hep-th/9704029.

\bi{dkm}
N. Dorey, V.V. Khoze  and M.P. Mattis, {\it Multi-Instantons, Three-Dimensional Gauge Theory, and the 
Gauss-Bonnet-Chern Theorem}, 
 \np   B502 (1997) 94, 
hep-th/9704197. 

\bi{bank}
 T. Banks, W. Fischler, N. Seiberg and L. Susskind, 
{\it Instantons, scale invariance 
and Lorentz invariance in matrix theory}, \pl B408 (1997) 111, hep-th/9705190.


\bi{bachas}
C. Bachas, {\it D-brane dynamics}, \pl B374 (1996) 37, hep-th/9511043.

\bi{kiri}
C. Bachas and E. Kiritsis, {\it $F^4$ terms in $N=4$ string vacua}, 
\np Proc. Suppl. B55 (1997) 194, 
hep-th/9611205.

\bibitem{ggr} O.J.~Ganor, R.~Gopakumar and S.~Ramgoolam, {\it Higher loop
effects in M(atrix) orbifolds}, hep-th/9705188.

\bi{kraus}  
 E. Keski-Vakkuri and  P. Kraus,
{\it Born-Infeld Actions from Matrix Theory},
hep-th/9709122.

    
\bi{dealwis}
S. P. de Alwis,  
{\it  Matrix Models and String World Sheet Duality},
hep-th/9710219.



\bi{ball}
V. Balasubramanian, R. Gopakumar and F. Larsen,
{\it Gauge Theory, Geometry and the Large N Limit},
hep-th/9712077.



\bibitem{bb} K.~Becker and M.~Becker, {\it A two loop 
test of M(atrix) theory}, \np B506 (1997) 48, hep-th/9705091. 

\bi{mald1}
J. Maldacena, {\it Probing near extremal black holes with D-branes}, hep-th/9705053.

\bi{yank}
W. Taylor, \np B508 (1997) 122, hep-th/9705116;
E. Keski-Vakkuri and P. Kraus, hep-th/9706196;
 J.M.~Pierre, hep-th/9707102;
 A. Brandhuber, N. Itzhaki, J. Sonnenschein and  S. Yankielowicz,
 hep-th/9711010.

\bi{tayl}
D.~Kabat and W.~Taylor,
{\it Spherical membranes in Matrix theory},
hep-th/9711078; Soo-Jong Rey, {\it Gravitating M(atrix) Q-Balls},
hep-th/9711081.


\bi{malda} J. Maldacena, {\it The large $N$ limit of superconformal field theories and supergravity}, hep-th/9711200.
\bi{rena}
R. Kallosh, J. Kumar and A. Rajaraman, {\it Special conformal symmetry of world volume actions}, hep-th/9712073.


\bi{bst}
E. Bergshoeff, E. Sezgin  and P.K. Townsend, 
 \pl B189 (1987) 75. 

\bi{town} 
M.J. Duff and J.X. Lu, \np B390 (19923) 276, hep-th/9207060;
P.K.~Townsend, 
%{\it D-branes from M-branes},
 Phys. Lett. B373 (1996) 68,
hep-th/9512062; 
C. Schmidhuber, \np B467 (1996) 146, hep-th/9601003;
A.A. Tseytlin, \np B469 (1996) 51, hep-th/9602064.  

\bi{dust}
M. Duff and K. Stelle, \pl B253 (1991) 113. 

\bi{RT}
J. Russo and A.A. Tseytlin, 
{\it Waves, boosted branes and BPS states in M-theory},
Nucl. Phys. B490 (1997) 121,  hep-th/9611047.

\bi{dual}
E. Bergshoeff, C. Hull and T. Ortin, \np B451 (1995) 547, hep-th/9504081;
J.C. Breckenridge, G. Michaud and R.C. Myers,
 Phys. Rev. D55  (1997) 6438,  hep-th/9611174.

\bi{witt} E. Witten, \np  B460 (1996) 335, hep-th/9510135.

\end{thebibliography}
\end{document}